\documentstyle[12pt]{article}

\input{psfig}
\begin{document}

\vspace*{-6ex}
\begin{flushright}
\fbox{FERMILAB-PUB-97/025-E}
\end{flushright}
\vglue 0.4cm

\begin{center}
{\large\bf Observation of Diffractive W-Boson Production at the Tevatron}
\vglue 0.6cm
{CDF Collaboration\\}
\vskip0.25in
\end{center}

\setlength{\baselineskip}{0.21in}

\begin{abstract}
We report the first observation of diffractively produced $W$ bosons.
In a sample of $W\rightarrow e\nu$ 
events produced in $p\bar{p}$ collisions at $\sqrt{s}$=1.8 TeV, 
we find an excess of events with a forward rapidity gap, which is
attributed to diffraction. The probability that this excess is consistent 
with non-diffractive production is $1.1\times 10^{-4}$ (3.8$\sigma$). 
The relatively low fraction of $W+Jet$ events observed 
within this excess implies that mainly quarks from the 
pomeron, which mediates diffraction, participate in $W$ production.
The diffractive to non-diffractive 
$W$ production ratio is found to be $R_W=(1.15\pm 0.55)\%$.
\end{abstract}

Approximately 15\% of high energy $p\bar p$ inelastic collisions 
are due to single diffraction dissociation, 
a process in which the incident $p$ or ${\bar p}$ 
escapes intact losing a fraction 
$\xi\leq 0.1$ of its initial forward momentum.
Experiments have shown \cite{G} that the leading role in 
diffraction is played by the pomeron \cite{Regge}, which carries 
the quantum numbers of the vacuum.  
In QCD the pomeron is a colorless entity, whose exchange in an event is 
marked by a ``rapidity gap", i.e. a large 
region of pseudorapidity \cite{pseudo} 
devoid of particles. 

The partonic structure of the pomeron was first investigated
by the UA8 experiment \cite{UA8_soft,UA8}, 
which studied diffractive dijet production at the 
CERN $S\bar p pS$ collider at $\sqrt{s}=630$ GeV,
and more recently by the H1 \cite{H1,H12} and ZEUS \cite{ZEUS1,ZEUS2} 
experiments in diffractive deep inelastic scattering (DDIS) 
\cite{H1,H12,ZEUS1} 
and dijet photoproduction \cite {ZEUS2} in $ep$ collisions at 
$\sqrt{s}\approx 300$ GeV at HERA. All experiments find that a substantial 
fraction of the pomeron structure is ``hard", i.e. consists of partons 
carrying  a large fraction of the pomeron momentum. 
From the DDIS experiments, which probe 
directly the quark component of the pomeron, 
the hard-quark component is estimated 
to account for about one third of the pomeron momentum.  
At the Tevatron
$\bar{p} p$ collider, a hard-quark 
pomeron structure 
would lead to detectable diffractive 
$W$ production \cite{BI}, which to leading order occurs through 
$q'\bar{q}\rightarrow W$. 
For a  hard-gluon dominated pomeron, $W$ production can occur through 
$qg\rightarrow Wq'$, but at 
a rate lower by order $\alpha_s$ and always in association with a jet. 

In this paper, we present the results of a measurement of  diffractive $W$ 
production in $p\bar p$ collisions at $\sqrt s=1.8$ TeV 
using the CDF detector at the Tevatron. Diffraction is tagged by the presence 
of a rapidity gap in an event in association with the following expected 
characteristic features.
In a diffractive $W^{\pm}\rightarrow e^{\pm}\nu$ event produced in 
a $\bar p$ collision with a pomeron ($\cal{P}$) emitted by the proton, 
the rapidity gap is expected to be at positive $\eta$ ($p$-direction)  
and the lepton boosted towards negative $\eta$ (angle-gap correlation). 
Also, since the pomeron is quark-flavor symmetric, 
and since from energy considerations mainly valence quarks from the 
$\bar p$ participate in producing the $W$, 
approximately twice as many electrons as positrons  
are expected (charge-gap correlation). These correlations can be seen in 
the Monte Carlo (MC) generated distributions of Fig.~1. 
The opposite correlations are, of course, 
expected for $p-{\cal{P}}$ collisions with the 
pomeron emitted by the $\bar p$.
In non-diffractive events, 
where rapidity gaps may arise from fluctuations in the event particle 
multiplicity, MC simulations
using the PYTHIA \cite{PYTHIA} program show that there are no significant 
angle-gap or charge-gap correlations.

We simulate diffractive events using the POMPYT \cite{POMPYT} 
MC program,  which is based on the Ingelman-Schlein 
model for hard diffraction \cite{IS}.
The  cross section for $p{\bar p}\rightarrow pX$ may be written as 
\[
\frac{d^2\sigma_{sd}^{p\bar p}}{dtd\xi}=
\left[K\;\xi^{1-2\alpha(t)}\;F^2(t)\right]\sigma_T^{{\cal{P}}{\bar p}}(\hat{s})
=f_{{\cal{P}}/p}(\xi,t)\;
\sigma_T^{{\cal{P}}{\bar p}}(\hat{s})
\]
where $K$ is a constant,  
$\xi$ is the fraction of the momentum of the proton carried by the pomeron,
$t$ is the square of the four-momentum transfer,
$\alpha(t)=1+\epsilon +\alpha't$ is the pomeron trajectory, 
$F(t)$ the nucleon form factor, 
$\hat{s}=\xi s$ the center of mass 
energy squared in the ${\cal{P}}-{\bar p}$ reference frame, 
and $\sigma_T^{{\cal{P}}{\bar p}}(\hat{s})$ the ${\cal{P}}-{\bar p}$ total 
cross section. 
This equation suggests the interpretation of single 
diffraction dissociation as a process in which a flux 
of pomerons, $f_{{\cal{P}}/p}(\xi,t)$, emitted by the proton interacts 
with the antiproton. This concept of factorization was extended \cite{IS} 
to hard processes by treating 
the ``pomeron flux factor" as a flux of particle-like 
pomerons with a unique partonic structure. In POMPYT, 
the collision of this flux of pomerons with the nucleon is
handled by PYTHIA. All our MC simulations include a simulation of the CDF 
detector.

The CDF detector is described in detail elsewhere \cite{CDF,CDF2}. 
In the rapidity gap 
analysis we use 
the ``beam-beam counters" (BBC) and the forward electromagnetic (EM) and 
hadronic (HA) calorimeters. 
The BBC \cite{CDF} consist of a square 
array of 16 scintillation counters on each 
side of the interaction point covering approximately 
the region $3.2<|\eta|<5.9$.
The forward calorimeters cover the region $2.4<|\eta|<4.2$ and 
have projective tower geometry 
with tower size 
$\Delta\eta\cdot \Delta\phi= 0.1\times 5^\circ$, where $\phi$ is the 
azimuthal angle. 
An energy threshold of 1.5 GeV (sum of EM plus HA energies) 
is used for each tower to exclude calorimeter noise.

The data sample was obtained during collider runs 1A (1992-1993) and 1B 
(1994-1995) by triggering on an electron of high 
transverse momentum, $P_T=P\sin \theta_e$, and on missing 
transverse energy, $\not\!\! E_T$ \cite{W}. 
We used events with $\not\!\! E_T>20$ GeV 
and an isolated \cite{isolation} electron of $E_T>20$ GeV 
in the central region, $|\eta|<1.1$, 
where the tracks of charged particles can be completely reconstructed. 
After implementing  a cut retaining events with one primary vertex only, 
8246 events remained. 
The one-vertex cut was imposed to exclude events with 
two interactions in the same beam-beam crossing, since 
the overlay of a ``minimum bias" on a diffractive $W$ event could 
eliminate the rapidity gap. 

We search for a diffractive $W$ signal by analyzing 
the correlations between the $\eta$ of the electron, $\eta_e$, 
or the sign of its charge, $C_e$, 
and the multiplicity of one or the other of the BBCs. Each event enters 
into two distributions, one with $\eta_e\cdot\eta_{BBC}<0$ (angle-correlated)
or $C_e\cdot\eta_{BBC}<0$ (charge-correlated), and the other with 
$\eta_e\cdot\eta_{BBC}>0$ (angle-anticorrelated) or 
$C_e\cdot\eta_{BBC}>0$ (charge-anticorrelated). 
A doubly-correlated (anticorrelated) distribution is the BBC multiplicity 
distribution for events with $\eta_e\cdot C_e>0$ and $\eta_e\cdot\eta_{BBC}<0$ 
($\eta_e\cdot \eta_{BBC}>0$). 
Fig.~2 shows the observed correlations as a function of 
BBC multiplicity, $N_{BBC}$, for events with tower 
multiplicity, $N_T$, 
less than 8 in the forward calorimeter adjacent to a given BBC. 
The cut on $N_T$ is imposed to reduce the non-diffractive 
contribution to the signal, since the signal is concentrated at low $N_{BBC}$ 
and is expected to have low $N_T$ 
as well. Fig.~2a shows the  angle and charge doubly-correlated (solid)
and doubly-anticorrelated (dashed) 
BBC multiplicities. The peaking at high multiplicities 
is caused by saturation due to the finite BBC segmentation. 
The two distributions agree well above 
the first three bins, but the correlated distribution  
has an excess in the first bin, 
consistent with the signature expected from diffractive events with a 
rapidity gap. This excess can be seen more clearly in Fig.~2b, 
which shows the bin-by-bin asymmetry (difference divided by sum) 
of the two distributions of Fig.~2a.
An excess is also seen 
in the individual angle (Fig.~2c) and charge (Fig.~2d) 
correlated asymmetries, as expected for diffractive production. 
From MC simulations of non-diffractive $W$ production and using 
Poisson statistics,
the probability that the observed excess in the first bin of
both the angle and charge correlated distributions is
due to simultaneous fluctuations in the non-diffractive
background was estimated to be $1.1\times 10^{-4}$.

The quark to gluon fraction of the 
partons of the pomeron participating in 
$W$ production may be evaluated
from the fraction of diffractive $W+Jet$ events observed.
Simulations performed with a hard-gluon (quark) pomeron structure predict the
fraction of diffractive $W$ events containing at least one jet with
$E_T>6$ GeV (within an $\eta-\phi$ cone radius of 0.7) to be 0.66 (0.20).  
For non-diffractive W events with similar kinematics
the predicted ``jet fraction" is 0.34, consistent with measurements in a
non-diffractive data sample.
In the first bin of Fig.~2a (solid histogram) there are 34 events, among
which we estimate 21 to be diffractive and 13 non-diffractive.  
Multiplying these numbers by the corresponding predicted jet fractions  
yields an expectation of $18.4\pm 2.8\, (8.8\pm 2.5)$ events with a jet 
for a hard-gluon (quark) pomeron structure.
The data contain 8 events with a jet, which implies that predominantly quarks
from the pomeron participate in W production.

We use the  
doubly-correlated distributions of Fig.~2a to evaluate the ratio, $R$, of 
diffractive to non-diffractive $W$ production 
rates. As a ratio, $R$ is insensitive to 
lepton selection cuts or inefficiencies and to the uncertainty in the 
luminosity. The acceptance for diffractive events is obtained from 
POMPYT using a hard-quark pomeron structure of the form 
$\beta G(\beta)=6\beta (1-\beta)$, where $\beta$ is the fraction of the 
momentum of the pomeron carried by the quark. In order to check for possible 
systematic effects due to BBC noise or inefficiencies that could distort 
the low multiplicity binning and thereby give an incorrect $R$ ratio, 
we evaluate $R$ using events with a BBC 
multiplicity upper bound, $N_B$, and we vary $N_B$  from 
zero to seven.  Fig.~3a shows the resulting $R$ values, 
and Fig.~3b the MC ``gap-acceptance", as a function of $N_B$. 
The gap-acceptance for bin $N_B$ is defined as the fraction of events with 
$N_{BBC}\leq N_B$ (the lepton acceptance is not included here).
The errors  in the points of Fig.~3a, which  are statistical, 
increase with increasing $N_B$ as more background is being subtracted.
To reduce the sensitivity of the result to the acceptance calculation, we 
retain as our signal the value $R=(1.03\pm 0.46)\%$ of the $N_B=2$ bin, where 
the acceptance is $81$\% and varies relatively slowly with $N_B$. 

As a systematic uncertainty in the gap-acceptance calculation we 
assign $\pm 13\%$, which is one half of the 
difference between the acceptances of
$N_B=1$ and $N_B=3$ divided by the acceptance of $N_B=2$. 
In deriving the ratio $R$ we assumed that the non-diffractive contributions
to the correlated and anticorrelated distributions in Fig.~2a are identical. 
This assumption is justified 
by the excellent matching of the two distributions for 
$N_{B}>3$.  A possible mismatch of the distributions 
within the available statistics introduces a systematic uncertainty, which 
was evaluated as follows. We made a straight line fit to 
the asymmetry of bins 4-10 of Fig.~2b, and extrapolated the fit  
into bins 1-3. For each of the bins 1-3,  
we multiplied the extrapolated asymmetry and its error by twice the number
of anticorrelated events, since the average number of non-diffractive 
correlated and anticorrelated events is expected to be the same, 
and added up the results for the three bins. 
Treating the sum as a signal yields a diffractive to non-diffractive ratio of 
$(0.01\pm 0.11)\%$, which is 
consistent with zero. We treat the error of
$\pm 0.11$\% as a systematic uncertainty in our measured value of $R$ and
add it in quadrature to the gap-acceptance uncertainty 
to obtain a combined systematic uncertainty of $\pm 0.18$\%.

From a study of the rate of W events versus instantaneous luminosity 
we estimate that a correction of $0.95\pm 0.05(syst)$ 
must be applied to $R$ due to the different efficiency 
of the single vertex cut for diffractive 
and non-diffractive events.
In addition, we apply a correction for the BBC occupancy by particles 
from a second interaction that does not have  a reconstructed
vertex.  From a study of a sample of 98000
events recorded by triggering the detector on beam-beam crossings only,
we determined that the probability of finding  
more than two hits in a BBC is $15\%$, corresponding to a BBC 
livetime acceptance 
of 0.85 by which we divide $R$.  The corrected value for $R$ is 
$R_W=[1.15\pm 0.51(stat)\pm 0.20(syst)]\%$. 
From MC simulations we estimate that 
the diffractive events are concentrated at 
$\xi$-values in the range 0.01-0.05.

Below we compare our results with POMPYT predictions and with results 
from other experiments. The predictions 
depend on the assumed pomeron structure function 
and on the form and normalization of the 
pomeron flux factor, $f_{{\cal{P}}/p}(\xi,t)$. We first use the 
``standard"  flux factor \cite{R} with 
parameters $\alpha(t)=1.115+0.26\,t$
and $K=0.73$ GeV$^{-2}$; for the nucleon form factor we use \cite{DL88}
$F(t)=({4m_p^2-2.8t})({4m_p^2-t})^{-1}\left[1-{t}/{0.7}\right]^{-2}$. 
For a two (three) flavor hard-quark 
pomeron structure of the form $\beta G(\beta)=6\beta(1-\beta)$ 
we obtain $R_W^{hq}$=24\% (16\%), while for a hard-gluon
structure of the same form, 
$R_W^{hg}=1.1\%$. Our measured ratio, $R_W=(1.15\pm 0.55)\%$, favors 
a purely gluonic pomeron, which however is incompatible with
the low fraction of diffractive $W+Jet$ events we observe. 
The HERA experiments on DDIS \cite{H1,ZEUS1} at $8.5<Q^2<65$ GeV$^2$ report a 
quark component in the pomeron structure which is flat in $\beta$, 
rises slowly with $Q^2$ at any given 
fixed $\beta$, and accounts for a fraction of 
about one third of the momentum 
of the pomeron, assuming the standard pomeron flux.
Independent of the pomeron flux normalization, by combining diffractive 
dijet photoproduction and DDIS results, the ZEUS collaboration reports 
\cite{ZEUS2} an integrated  hard-quark momentum fraction of $0.2<f_q<0.7$,
while the H1 collaboration \cite{H12}, from a QCD analysis of DDIS, 
obtains $f_q\approx 0.2$ at $Q^2\sim 60$ GeV$^2$. 
The $Q^2$ evolution from $Q^2=60$ GeV$^2$ to 
$Q^2=M_W^2$ of the pomeron structure function proposed by H1  
does not change significantly 
the quark component participating in $W$ production.
Using a pomeron with a hard-quark fraction of 0.2 and a gluon fraction of 0.8, 
POMPYT predicts ratios $R_W$ of 5.7\% (4.1\%) 
for two (three) quark flavors, which are larger than our measured 
value of ($1.15\pm0.55$)\% by more than eight (five) standard deviations.

We now compare our results with POMPYT predictions 
using the ``renormalized" pomeron flux \cite{R}, 
defined as the standard flux normalized, if its integral exceeds unity,
to one pomeron per nucleon. 
The normalization factor is $\approx 9$ 
at $\sqrt{s}=1.8$ TeV (CDF) and $\approx 1$  at HERA (see \cite{R}).
The predictions for $R_W$ become 2.7\% (1.8\%) for a two (three) flavor
pure hard-quark and 0.12\% for a pure hard-gluon 
pomeron structure. Based on these 
predictions, our $R_W$ value of ($1.15\pm0.55$)\% implies 
hard-quark fractions of 
$f_q=0.4\pm 0.2$ $(0.6\pm 0.3)$ for two (three) quark flavors. 
These fractions are consistent with the ZEUS and H1 results of 
$0.2<f_q<0.7$ and $f_q\approx 0.2$, respectively. 
Assuming the momentum sum rule, $f_q+f_g=1$, 
the predicted fractional gluon contribution to $R_W$ is 
$(0.12\%)(1-f_q)/[(0.12\%)(1-f_q)+Af_q]$, where $A$=2.7\% (1.8\%) 
for two (three) quark flavors.  
From our values of $f_q$, the gluon contribution to $R_W$
is predicted to be 6.6\% (4.2\%) for two (three) quark flavors, which can 
explain the low fraction of $W+Jet$ events we observe.

In conclusion, we have observed diffractive $W$ production in $p\bar p$ 
collisions at $\sqrt{s}=1.8$ TeV and measured the ratio of 
diffractive to non-diffractive production rates to be 
$R_W=(1.15 \pm 0.55)\%$. The   
relatively small fraction of diffractive $W+Jet$ events we observe implies 
that mainly quarks from the pomeron participate in diffractive $W$ production.

We thank the Fermilab staff and the technical staffs of the
participating institutions for their vital contributions.  This work was
supported by the U.S. Department of Energy and National Science Foundation;
the Italian Istituto Nazionale di Fisica Nucleare; the Ministry of
Education,Science and Culture of Japan; the Natural Sciences and
Engineering Research Council of Canada; the National Science Council
of the Republic of China; and the A. P. Sloan Foundation.

\newpage
\begin{figure}[htbp]
\vspace*{1.5in}
{\hspace*{0.9in}
{\psfig{figure=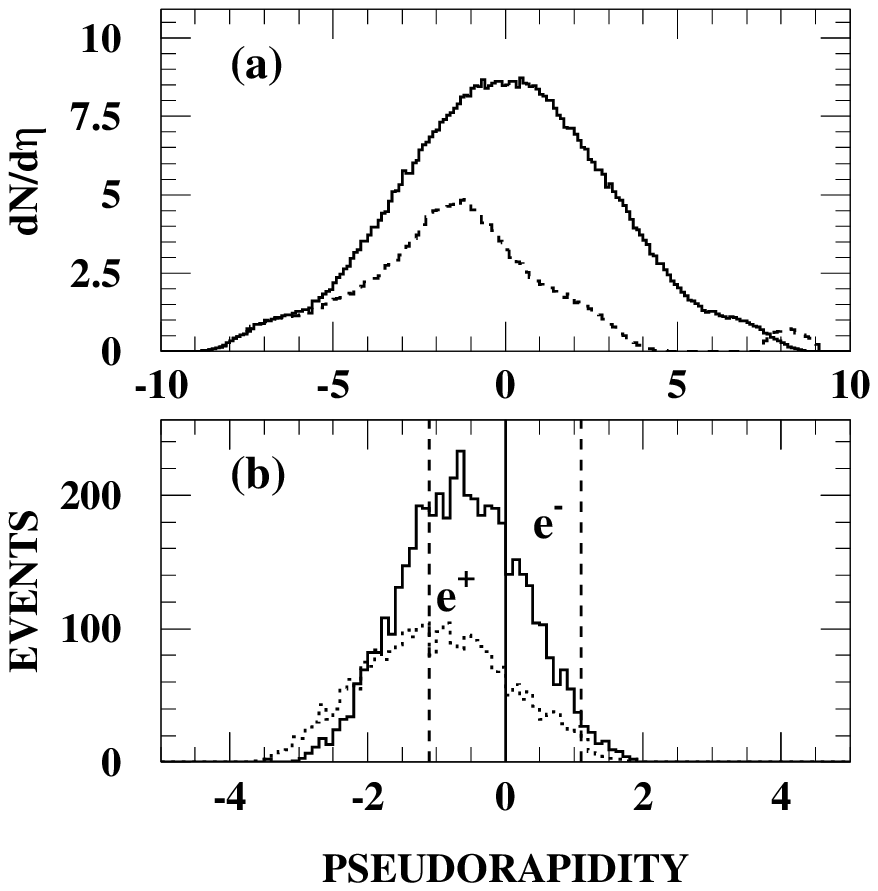}}}
\vspace*{-3.25in}
\caption{ Monte Carlo generated $\eta$-distributions: $(a)$
Particle densities for non-diffractive
(solid) and for diffractive (dashed) $W$ events for
pomerons of beam momentum fraction 
$\xi =0.03$ emitted by protons (at positive $\eta$);
the small bump at $\eta\approx 8.5$ is caused by the leading protons.
$(b)$ Electrons and positrons
from diffractive $W^{\pm}(\rightarrow e^{\pm}\nu)$ 
events for all pomerons of $\xi<0.1$
emitted by protons (the vertical dashed lines define the boundaries
of the region of this measurement).}
\label{W2}
\end{figure}
\clearpage
\newpage
\begin{figure}[htbp]
{\hspace*{1in}
{\psfig{figure=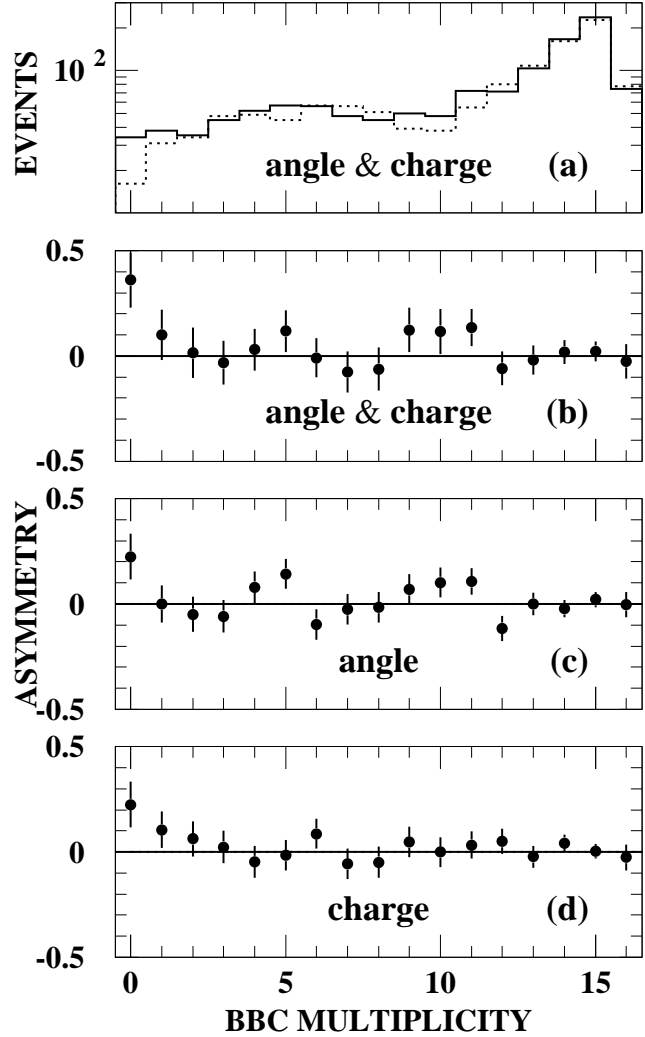}}}
\vspace*{-1.75in}
\caption{$(a)$ Electron angle and charge doubly-correlated (solid) 
and anticorrelated (dashed) distributions (see text) 
versus BBC multiplicity, and
$(b)$ the corresponding asymmetry, defined 
as the bin-by-bin difference over sum of the two 
distributions in $(a)$. 
The diffractive signal is seen in the first bin as an 
excess of events in the 
correlated distribution in $(a)$, and 
as a positive asymmetry in $(b)$. An asymmetry is also seen in the first bin
of the individual angle $(c)$ and charge $(d)$ distributions.}
\label{W3}
\end{figure}
\clearpage
\newpage
\begin{figure}[htbp]
\vspace*{1in}
{\hspace*{1in}
{\psfig{figure=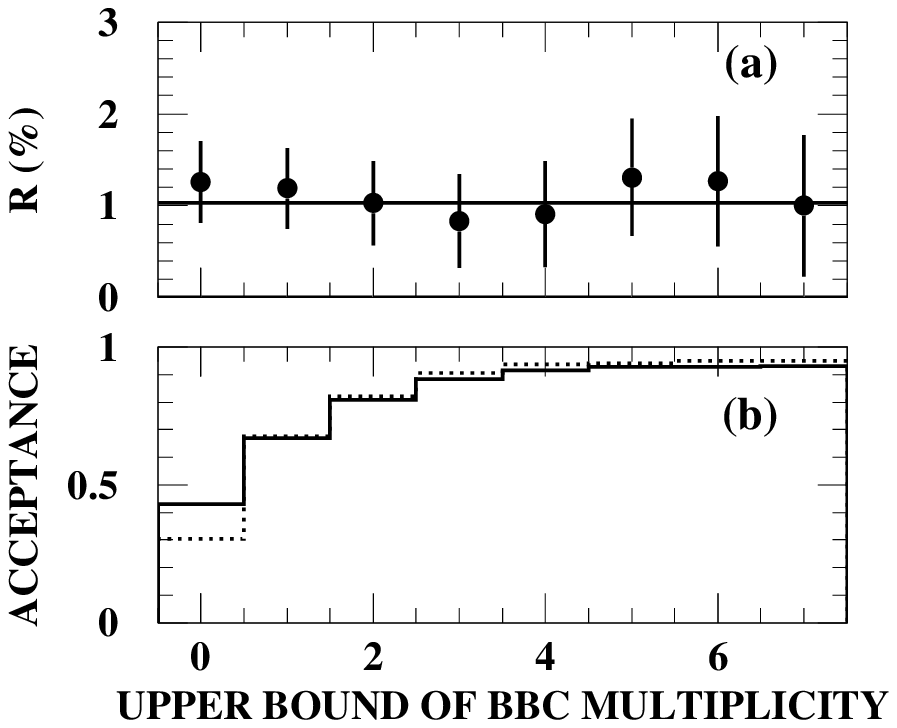}}}
\vspace*{-4in}
\caption{$(a)$ Diffractive to non-diffractive $W$ production ratio  
(not corrected for BBC occupancy or one-vertex cut efficiency) as a function 
of upper bound $BBC$ multiplicity, $N_B$. 
The solid line passes through the $N_B=2$ point, which
we use as our result;
$(b)$ gap-acceptance 
for angle-gap and charge-gap doubly-correlated (solid) and 
anticorrelated (dashed) 
diffractive events with an electron within $|\eta|<1.1$.
}
\label{W4}
\end{figure}


\newpage
\begin{thebibliography}{99}
\bibitem {G} K. Goulianos, Phys. Reports {\bf 101}, 169 (1983).
\bibitem{Regge} P.D.B. Collins, An Introduction to Regge Theory and High 
Energy Physics, Cambridge University  Press, Cambridge (1977).
\bibitem{pseudo}We use rapidity and pseudorapidity, $\eta$, interchangeably;
$\eta\equiv -\ln(\mbox{tan}\frac{\theta}{2})$, 
where $\theta$ is the polar angle of
a particle with respect to the proton beam direction.
\bibitem{UA8_soft}R. Bonino {\em et al.}, Phys. Lett. {\bf B 211}, 239 (1988).
\bibitem{UA8} A. Brandt {\em et al.}, Phys. Lett. {\bf B 297}, 417 (1992).
\bibitem{H1} T. Ahmed {\em et al.}, 
Phys. Lett. {\bf B 348}, 681 (1995).
\bibitem{H12} H1 Collaboration, A Measurement and QCD Analysis of the 
Diffractive Structure Function $F_2^{D(3)}$, 
submitted to ICHEP'96, Warsaw, Poland, July 1996. 
\bibitem{ZEUS1}M. Derrick {\em et al.}, 
Z. Phys. {\bf C68}, 569 (1995).
\bibitem{ZEUS2} M. Derrick {\em et al.},
Phys. Lett. {\bf B 356}, 129 (1995).
\bibitem{BI} P. Bruni and G. Ingelman, Phys. Lett. {\bf B 311}, 317 (1993).
\bibitem{PYTHIA} T. Sj\"{o}strand, Comput. Phys. Commun. {\bf 82}, 74 (1994).
\bibitem{POMPYT} P. Bruni and G. Ingelman, 
Preprint DESY-93-187; Proceedings of the International Europhysics Conference 
on High Energy Physics, Marseille, France, 22-28 July 1993, Editions
Fronti\`{e}res (Eds. J. Carr and M. Perrottet) p.595.
\bibitem{IS} G. Ingelman and P. Schlein, Phys. Lett. {\bf B 152}, 256 (1985).
\bibitem{CDF}F.Abe {\em et al.}, 
Nucl. Instrum. Methods {\bf A 271}, 387 (1988). 
\bibitem{CDF2}D. Amidei {\em et al.}, Nucl. Instrum. Methods  {\bf A 350},
73 (1994).
\bibitem{W} The transverse energy, $E_T$, measured by 
a calorimeter cell located at polar angle $\theta$ is defined as $E\sin\theta$. 
Missing $E_T$, $\not\!\! E_T$, 
is defined as the magnitude of the vector that balances  
the vector sum of $E_T$ in all calorimeter cells within $|\eta|<3.6$.
\bibitem{isolation}The calorimetric isolation, $I_{cal}$, is defined as the 
sum-$E_T$ in the towers within a cone of radius 
$r=\sqrt{(\Delta\eta)^2+(\Delta\phi)^2}=0.4$ centered on the electron, 
excluding the $E_T$ of the electron. The ``isolated electron" requirement is 
$I_{cal}/E_T^e<0.1$.
\bibitem{R} K. Goulianos, Phys. Lett. {\bf B 358}, 379 (1995).
\bibitem{DL88} A. Donnachie and P. V. Landshoff, Nucl. Phys. {\bf B 303},
634 (1988).
\end{thebibliography}
\end{document}